\documentclass[sigconf]{acmart}

\usepackage{graphicx}
\usepackage{threeparttable}
\usepackage{framed}
\usepackage{enumitem}
\usepackage{todonotes}
\usepackage[skip=3pt]{caption} 

\title{Teaching Prompt-Based Programming with LLMs: A 45-Minute Lesson with Guided Practice for End-User Programmers}

\author{Keith Tran}
\email{ktran24@ncsu.edu}
\affiliation{%
  \institution{North Carolina State University}
  \city{Raleigh}
  \state{North Carolina}
  \country{USA}
}

\author{Samiha Marwan}
\email{jmskripc@ncsu.edu}
\affiliation{%
  \institution{North Carolina State University}
  \city{Raleigh}
  \state{North Carolina}
  \country{USA}
}

\author{Thomas Price}
\email{tprice@ncsu.edu}
\affiliation{%
  \institution{North Carolina State University}
  \city{Raleigh}
  \state{North Carolina}
  \country{USA}
}

\copyrightyear{202X}
\acmYear{202X}
\setcopyright{acmlicensed}
\acmConference[Conference acronym 'XX]{Make sure to enter the correct
  conference title from your rights confirmation email}{June 03--05,
  2018}{Woodstock, NY}
\acmDOI{XXXXXXX.XXXXXXX}
\acmISBN{978-1-4503-9976-0/23/08}

\begin{document}

\begin{abstract}

Prompt-based programming, a new modality enabled by large language models (LLMs), allows users to express computational goals through natural language rather than traditional code. While this approach lowers barriers to entry, especially for non-CS learners, it does not eliminate the need for foundational CS skills. Learners often struggle to communicate their intent clearly to LLMs, resulting in vague or underspecified prompts. Prior work has documented the need for explicit prompting for both CS and non-CS learners. However, it remains less clear how such instruction can fit into busy classrooms or how much time is needed to produce meaningful gains. In this paper, we evaluated a 45-minute prompt-based programming intervention, consisting of a lesson with guided practice, against a business-as-usual CS lab activity (code tracing) of equal length, representing a class without prompt-focused instruction. We conducted a randomized controlled study with 55 engineering students. We found that students in the experimental condition improved more on average (though not significantly more) from pre- to post-test than the control group (+10.8 vs +1.1 percentage points) and showed significantly greater average gains in prompting self-efficacy (+35.4 vs +21.9 percentage points). Our results suggest it is likely that a brief intervention can improve learners' ability to specify computational goals to LLMs. However, the effect was modest, suggesting that prompting skills may require more time and practice to develop. We provide a lightweight lesson that requires no prior CS background and can be readily dropped into existing courses.

\end{abstract}


\maketitle

\section{Introduction}
\label{sec:intro}

Recent advances in large language models (LLMs), such as ChatGPT, have introduced a new way for learners, especially non-CS-majors, to engage with programming: describing a task in everyday language (e.g., “take a list of numbers and filter out those between 1 and 10”) and receiving working code in return. This practice, called prompt-based programming, allows users to express computational intent without writing traditional code \cite{jiang2022promptmaker}. However, access to LLMs alone is not sufficient: research has shown that users, especially those with less programming experience, often struggle to communicate their intent clearly. Common barriers include using vague or underspecified prompts \cite{nguyen2024how, liange2025prompts, feldman2024nonexpert, denny2023prompt}, lacking strategies for debugging or refinement \cite{pereira2023why, nguyen2024how}, and lacking technical CS vocabulary \cite{feldman2024nonexpert}. These findings raise the question of whether targeted instruction can help learners better articulate computational goals and debug prompts to generate correct solutions.

Such prompting-specific programming lessons may be particularly impactful for learners with \textit{end-user programming} goals, who want to learn to write code to accomplish domain-specific objectives, ``as a means to an end,'' rather than ``the end itself,'' \cite{ko2011state, sarkar2023will}. While there is a clear need to teach prompting for both CS and non-CS learners, it remains less clear how such instruction can fit into busy classrooms or how much instructional time is needed to produce meaningful gains. This question is especially valuable for instructors who see the benefits of teaching prompting but are unsure how to integrate it into their curricula.

To explore these questions, we designed a 45-minute lesson on prompt-based programming paired with guided practice. We evaluated this lesson plus practice intervention against a business-as-usual CS lab activity (code tracing) of equal length, representing a class without prompt-focused instruction. 


We conducted a randomized controlled study with 55 engineering students, comparing these two lessons during a single lab session. To assess students' prompting ability, we used Promptly problems \cite{denny2023promptly} as pre- and post-tests. These tasks require learners to identify patterns in input/output pairs (i.e., test cases) and prompt an LLM to write a function that passes the test cases, \textit{without} access to a natural language problem description that could be copied into the LLM. Our study is guided by two research questions: What is the impact of a short, prompt-based instruction on RQ1) learners’ ability to successfully solve Promptly problems and RQ2) learners' self-efficacy in prompting skills? This paper makes the following contributions:

\begin{itemize}[nolistsep, topsep=1.5pt]
    \item We present a lightweight, adaptable 45- minute lesson with guided practice that introduces CS-specific prompting strategies for use with LLMs. All instructional materials are publicly available\footnote{URL omitted for review to preserve anonymity.}.
    \item We demonstrate through a controlled experiment that this short intervention can significantly increase learners' prompting self-efficacy over a business-as-usual CS lab activity, with meaningful (though not statistically significant) improvements in their ability to solve Promptly problems.
\end{itemize}

%
%
\section{Related Work}



\textit{\textbf{Prompt-Based Problems in CS Education}}: Prompt-based programming has emerged as a new paradigm in response to the rise of LLMs, allowing learners to express computational intent through natural language rather than code. To support this shift, Denny et al.\ introduced \textit{Prompt Problems} and the \textit{Promptly} platform, framing prompt construction as a teachable skill and providing automated feedback on LLM-generated code \cite{denny2023prompt, denny2023promptly, kerslake2024intergrating}. These problems assess students’ ability to express intent clearly, rather than reproduce syntax, and are designed to resist shortcut solutions like copy-pasting. However, recent work shows that novice learners often struggle with these tasks \cite{kazemitabaar2024how}. Nguyen et al.\ found that CS1 students solved just 57\% of prompt problems, typically after multiple attempts and with automated feedback \cite{nguyen2024how}. Students relied on ad hoc refinement strategies and held incorrect mental models of how code-generating LLMs work. Feldman and Anderson observed even lower success among non-experts, citing vague prompts and limited CS vocabulary as key barriers \cite{feldman2024nonexpert}, and recent work suggests that information content, such as what students include, matters more than vocabulary alone \cite{lucchetti2025substance}. Together, these findings suggest that while LLMs lower barriers to entry, non-programmers face persistent challenges in expressing computational goals. This barrier is likely shared by conversational programmers, who engage with code conceptually rather than as authors \cite{wang2018mismatch, hur2024profiling}. These challenges highlight the need for explicit instruction in prompt-based programming.

\textit{\textbf{Prompt-Based Programming Instruction}}:  
A growing body of work explores how prompting can be taught explicitly. Prather et al.\ provide a comprehensive review of GenAI research and teaching practices in computing education, noting that new tools are emerging to teach both programming and prompting skills simultaneously \cite{prather2025beyond}. While best practices and frameworks—such as OpenAI guidance or the CLEAR framework (Concise, Logical, Explicit, Adaptive, Reflective) \cite{lo2023clear}-offer general advice, they often do not transfer well to domain-specific tasks like programming. Liang et al.\ highlight this challenge in their study of prompt developers, observing that prompt programming differs from traditional software development. Developers often lack reliable mental models of LLM behavior, and prompts are shaped through iterative trial-and-error, rather than systematic design \cite{liange2025prompts}. They conclude that prompt programming lacks standardized best practices and recommend instructional approaches that are tailored to this emerging skill.

One promising example is the ROPE intervention, which teaches learners to articulate task requirements for code generation \cite{ma2025rope}. ROPE and our work share the goal of making prompting more systematic and teachable. However, ROPE is designed to support general-purpose prompt engineering, whereas our lesson focuses on foundational CS instruction—teaching novices to express computational goals through prompts aligned with core programming constructs. Both efforts reflect a broader need for more research on how to teach prompt-based programming effectively, especially for learners entering computing through LLMs.

\textit{\textbf{Theoretical Framework}}:  
The design of our intervention is guided by the Gulf of Envisioning framework \cite{subramonyam2024bridging}, which helps explain why learners may fail even when using powerful AI tools. By scaffolding learners’ ability to describe the problem (\textit{capability}), phrase prompts effectively (\textit{instruction}), and assess output quality (\textit{intentionality}), we aim to reduce the cognitive gap between intent and implementation. Our lesson incorporates explicit instruction and practice for each of these dimensions, which helps frames prompt-based programming as an alternative CS-specific skill that can be learned and improved through guided support.

\vspace{-4mm}
\section{Intervention}
\label{sec:intervention}

This study investigates whether a short prompt-based lesson could help end-user programmers communicate computational intent more effectively to large language models (LLMs). We developed an experimental lesson that integrates selected CS1/CS2 concepts (e.g., function structure, data transformation, and edge cases) with prompt engineering strategies tailored for LLM interactions.

\textit{\textbf{Learning Objectives}}: Our lesson was developed using a backward design approach \cite{wiggins2005understanding}, beginning with the end goal: enabling learners to construct effective prompts for LLMs to solve Promptly-style programming tasks. These tasks differ from traditional programming tasks in that they require specifying computational intent in natural language, rather than writing code directly. To achieve this goal, we selected two representative problem types commonly taught in CS1: data transformation and conditional logic. These task types reflect computational patterns (detailed below) in early programming courses, while remaining accessible and appropriate for our target population of end-user programmers \cite{pereira2023why}.

Our objectives were also informed by challenges observed in prior studies. Feldman et al. \cite{feldman2024nonexpert} found that novice LLM users lacked CS vocabulary, conflated technical and colloquial terms, struggled to revise prompts after failures, and often produced underspecified prompts. To address these challenges, we identified five core learning objectives, aligned with the three gaps in the Gulf of Envisioning framework. By the end of the intervention, students should be able to:
\begin{itemize}[nolistsep, topsep=1.5pt]
    \item \textbf{Specify function structure}: Define the input, outputs, and what the function does. This foundational skill helps learners overcome the \textit{instructional gap} by translating goals into specifications that an LLM can interpret.
    \item \textbf{Describe data transformation}: Describe how the input should be changed into the output. We emphasized three common patterns: (1) mapping, (2) filtering, and (3) reducing, which reflect key capabilities of LLM-generated code and help address the \textit{capability gap}.
    \item \textbf{Express conditional logic}: Use branching constructs (e.g., \texttt{if}/\texttt{else}) to describe decision-making behavior.
    \item \textbf{Handle edge cases}: Reason about special or boundary inputs (e.g., empty strings, zero-length lists) to ensure prompt correctness across varying inputs.
    \item \textbf{Debug failed prompts}: Interpret outputs and iteratively refine prompts to better align with intent, helping students close the \textit{intentionality gap}. 
\end{itemize}

\textit{\textbf{Lesson Structure}}: 
The lesson was designed to last 45 minutes to fit within a single class period and focused exclusively on prompting skills, with no instruction in Python syntax or code comprehension. While short for an intervention, this duration reflects a realistic constraint: if prompting can be meaningfully taught in one session, it could be integrated into existing courses without substantial disruption. It consisted of 5 short sections (\~5 minutes each), one for each learning objective above, delivered via lecture and slides. After each section, students engaged in active learning activities, including live demonstrations and 3 Promptly problems with automated feedback, which gave students the opportunity to practice what they learned in the lesson. These practice problems targeted the same learning objectives as the pre/post-test (e.g., filtering, conditional logic, edge cases) but used different problem contexts to assess transfer. For example, students practiced filtering words by length, while the corresponding test item required filtering numbers by range. As described in Section~\ref{sec:methodology}, our experiment compared this experimental prompting lesson to a control lesson with the same duration, structure, and number of practice opportunities.

To help students remember to address each learning objective, the lesson taught students a reusable template to construct prompts, which were built up throughout the lessons:

\vspace{-0.1\baselineskip}
\begin{framed}
 Write me a Python function that takes as input [\textit{describe the input, including its type}] and it should output [\textit{describe the output, including its type}], that [\textit{explain how the function transforms the input into the output}]. \newline For example: [\textit{provide an example input and expected output}].
\end{framed}

\vspace{-4mm}
\section{Methodology}

\label{sec:methodology}
We conducted a randomized controlled experiment with a pre-test/post-test design,evaluating a 45-minute prompt-based programming lesson with guided practice against a business-as-usual CS lab session of equal length.

\textit{\textbf{Recruitment}}: After receiving IRB approval, we recruited participants from a first-year engineering course at Anonymized University, by asking the course instructor to send emails to students announcing our research study. This course is required for all engineering students prior to declaring their major, making it an ideal context for studying end-user programmers, students who are likely to engage with programming in their coursework or professional career, but who do not necessarily intend to major in CS.

To focus on students with limited formal CS experience, we recruited students from the second 8-week section of the course. The university’s enrollment policy places students co-enrolled in a programming course into the first 8-week session, while those not concurrently enrolled in programming typically enter the second. While some students reported prior experience through high school, college coursework, or informal learning, most were not enrolled in a formal programming course during the study. Recruitment materials informed all students that the activity would involve prompting an LLM to solve programming tasks, but described the learning session generically as developing "programming skills" without specifying the instructional approach or the existence of multiple conditions. 

A total of 58 students consented and completed the study. Participants included 36 men and 22 women, with the majority (79\%) in their first year. The sample's racial composition: 48\% White, 19\% Asian, 14\% Black or African American, 12\% Hispanic/Latino, and 6\% other or preferred not to answer. Sixteen percent were non-native English speakers, though all reported English proficiency. Intended majors spanned engineering disciplines, including mechanical (12\%), electrical (7\%), civil (7\%), and chemical engineering (5\%), with 7\% in exploratory studies and the remainder distributed across other technical fields.

\textit{\textbf{Control Condition}}:
Our control condition is a 45-minute lesson on code tracing, a common topic in introductory programming. We chose code tracing to represent what a business-as-usual programming lab session, \textit{not} focused on teaching prompting, might cover in a 45 minute lesson. Since the control condition does not teach or practice prompting, this allows us to measure the effect of our intervention and what benefits it might offer if integrated into a classroom. The control condition was designed to be structurally parallel to the Experimental (see Table~\ref{tab:conditions}). Both lessons used the same 45-minute format, with approximately 25 minutes of direct instruction followed by 20 minutes of active practice. Both addressed the same computational concepts—filtering, conditionals, and edge cases—and included 3 practice problems. The key difference was instructional focus: while the experimental lesson emphasized prompt structure, data transformation patterns, and debugging prompts, the control lesson introduced basic Python syntax, control flow, and code tracing. Practice activities also differed in format: the experimental group completed prompt-writing tasks on Promptly, while the control group worked through ``explain-in-plain-English'' tracing exercises using Python Tutor \cite{guo2013}. Because the control centered on code tracing, practice activities used code-tracing exercises rather than prompt-writing tasks. Python comprehension may still help students interpret LLM-generated code on Promptly, so any gains we observed in the experimental condition reflect the added value of prompting instruction beyond what business-as-usual instruction would provide.

\textit{\textbf{Conditional Assignment}}: Randomization occurred at the session level. Students signed up for time slots before randomization and were unaware of which condition each session would receive. Sessions were paired by enrollment size, then randomly assigned to condition.

\textit{\textbf{Session Delivery}}: To accommodate scheduling needs, we offered both in-person and online sessions, all delivered synchronously by the first author. We conducted a total of 15 sessions. The experimental condition included 7 sessions (M = 4.29 participants/session; SD = 2.21), and the control condition included 8 sessions (M = 3.62 participants/session; SD = 2.50). In total, there were 30 participants in the Experimental group and 28 in the Control group. Each session followed a single experimental condition.

\textit{\textbf{Procedure:}} Each session lasted 90 minutes and followed a fixed structure across both conditions (see Table~\ref{tab:conditions}): 
(1) \textit{Pre-survey} (5 min): Assessed prior programming experience and baseline self-efficacy. (2) \textit{Pre-test} (15 min): Students completed 8 Promptly problems (see Section~\ref{sec:intro} for details). (3) \textit{Instructional intervention} (45 min): Students received either the prompt-based or code-comprehension lesson (see Control Condition above for details) (4) \textit{Post-test} (15 min): Students completed a parallel version of the Promptly assessment (see Assessments below). (5) \textit{Post-survey} (5 min): Assessed post-intervention prompting self-efficacy and collected feedback.


\begin{table}[t]
    \centering
    \footnotesize
    \begin{threeparttable}
    \caption{Comparison of Experimental and Control Conditions}
    \label{tab:conditions}
    \renewcommand{\arraystretch}{1.3}
    \begin{tabular}{@{}p{1.8cm}p{2.8cm}p{2.5cm}@{}}
        \toprule
        & \textbf{Experimental} & \textbf{Control} \\
        \midrule
        \textbf{Lesson Content} &
        Prompt structure, data transformation, debugging &
        Python syntax, control flow, code tracing \\
        \addlinespace
        \textbf{Target Skill} &
          Writing effective prompts &
          Reading/understanding code \\
        \addlinespace
        \textbf{Practice Format} &
        Prompt-writing exercises on Promptly (×3) &
        Code-tracing exercises on Python Tutor (×3) \\
        \addlinespace
        \textbf{Practice Topics} &
          \multicolumn{2}{c}{Filtering,  Conditionals, Edge cases} \\
        \addlinespace
        \textbf{Feedback} &
        LLM-generated code &
        Step-by-step visualization \\
        \bottomrule
    \end{tabular}
    
    \begin{tablenotes}
        \footnotesize
        \item Note: Both conditions used the same duration (45 min)
    \end{tablenotes}
    \end{threeparttable}
    \vspace{-4mm}
\end{table}

\textit{\textbf{Measures}}:
\textit{Prior Programming Experience}. To measure students' prior programming experience, we included pre-survey questions asking students to indicate: 1) the number of courses they had previously taken involving programming, 2) how frequently they programmed on a 6-point ordinal scale, and 3) their self-assessed comfort completing hypothetical programming tasks of varying difficulty (e.g., from printing "Hello, World!" to building a small application) on a 7-point ordinal scale. While there is no accepted standard for assessing prior programming experience, these questions concisely captured the major aspects of prior experience that are measured in prior work: prior courses \cite{bui2023prior, duran2019towards}, frequency \cite{duran2019towards}, and self-assessed comfort \cite{ feigenspan2012measuring}. We constructed a composite index of these 3 measures: we first z-score normalized each response, so that each measure used the same units of standard deviations from the mean, and then averaged these z-scores across the 3 measures\footnote{As discussed later, our results were robust to different operationalization of prior experience, e.g. using just one of these measures, or including them as separate model variables, so we present the average for simplicity}.

\textit{Prompting Self-efficacy}. We adapted three items from the General LLM AI Self-Efficacy Scale \cite{ju2025developing} to measure students’ confidence in using LLMs for programming tasks. Traditional programming self-efficacy scales do not reflect the unique challenges of prompt-based programming, so we tailored items to this context. Students rated their confidence (0 = Absolutely No Confidence to 10 = Extremely Confident) on three items: (1) providing effective prompts to LLMs, (2) judging the correctness of LLM outputs, and (3) completing programming tasks using LLM-based tools.

\textit{\textbf{Assessments}}:
Assessments were created to align with the lesson content and reflect the learning objectives emphasized during instruction (see Section~\ref{sec:intervention}). While we initially reviewed existing assessments (e.g., \cite{nguyen2024how, denny2023promptly}), we found they did not align with our instructional content or target population of end-users. As a result, we developed our own items to better reflect the core skills emphasized in the intervention and support better comparison between conditions.

Each assessment included eight Promptly problems, with task types including transformations (e.g., reversing a word), filtering (e.g., removing non-positive numbers), conditional logic, aggregation (e.g., computing an average), and combinations of these. We also included a mapping task (e.g., converting [1, 2, 3] to "abc"). To ensure breadth, we varied input/output types across tasks, including words, numbers, and lists. For example, one pre-test task asked to filter non-positive numbers from a list, while the corresponding post-test task asked to filter numbers greater than 10. The full set of problems is available at [anonymized link].

\textit{\textbf{Data Analysis}}:
To evaluate the impact of our instructional intervention on student performance (RQ1) and prompting self-efficacy (RQ2), we used linear mixed-effects models to account for repeated measures and session-level clustering \cite{winter2013linear}. Each model included fixed effects for Time (pre vs. post), Condition (Control vs. Experimental), and their interaction. We report unstandardized regression coefficients (B) and 95\% confidence intervals as effect size measures, representing expected change in problems solved (out of 8) for RQ1 and points on a 0–10 scale for RQ2. This approach follows recommendations for reporting effect sizes in regression frameworks \cite{baguley2009standardized}, where unstandardized coefficients provide interpretable estimates of practical significance, while controlling for random effects.

We also included two standardized covariates, Prior Programming Experience and Prompting Self-Efficacy, to control for baseline individual differences that might influence student outcomes. For the self-efficacy outcome (RQ2), we removed self-efficacy as a covariate and included the performance of the pretest instead. The model assumptions were verified through residual analysis. Before fitting models, we excluded three participants from analysis: one who achieved a perfect pre-test score, one who did not complete both assessments, and one who bypassed the task by submitting working code directly instead of writing a prompt.

We modeled outcomes at the student level and included random intercepts for both Student and Session to account for repeated measures and potential clustering. Intra-class correlation coefficients (ICCs) from the fitted models (Tables~\ref{tab:task-performance} and \ref{tab:self-efficacy}) for pre-test performance (ICC = 0.013) and prompting self-efficacy (ICC $\approx$ 0.000) suggest that the session in which a student completed the study was not meaningfully associated with their outcomes.

\vspace{-1mm}

\section{Results}

We analyzed data from 55 participants (28 experimental, 27 control) who completed a randomized, session-controlled study comparing prompt-based and code-comprehension instruction. All participants completed pre- and post-tests on Promptly-style tasks and surveys measuring prior programming experience and prompting self-efficacy. Baseline measures were balanced across groups. Prior programming experience was similar (Control: M = –0.084, SD = 0.726; Experimental: M = 0.068, SD = 0.945), as was pre-test performance (Control: M = 3.63, SD = 1.50; Experimental: M = 3.71, SD = 1.57). Prompting self-efficacy was slightly higher in the experimental group (Control: M = 2.63, SD = 2.58; Experimental: M = 3.26, SD = 2.40), though this difference was not significant and is controlled for in the models (see Section~\ref{sec:rq2}). 

\textit{\textbf{Approach to Significance Testing}}: Based on modern recommendations from the American Statistical Association \cite{wasserstein2016asa} and others \cite{wasserstein2019moving} we limit this paper's emphasis on statistical significance at $\alpha = 0.05$, treating it as a measure of a model's fit to data, rather than a dichotomous indicator of whether a result is important or not. Instead, we emphasize effect sizes (in this, case model beta-weights), confidence intervals (suggesting the likely range of values of that effect size), and practical significance, and we acknowledge the uncertainty in all effects that we report.
We now present results addressing our two research questions: RQ1, the impact of prompt-based instruction on task performance, and RQ2, its impact on prompting self-efficacy.



\subsection{RQ1: Task performance}

We used a linear mixed-effects model to analyze students' programming task performance, measured as the number of problems solved (out of 8), with partial credit awarded based on the proportion of test cases passed per problem. The model included fixed effects with covariates, and random intercepts as described in the \textit{Data Analysis} section. Model assumptions were met: residuals were approximately normally distributed (Shapiro–Wilk $W \approx$ 0.98, $p$ = 0.81), and residual plots showed no evidence of non-linearity or heteroscedasticity.

\begin{figure}[ht]
  \centering
   \vspace{-2.5mm} 

\includegraphics[width=0.85\linewidth]{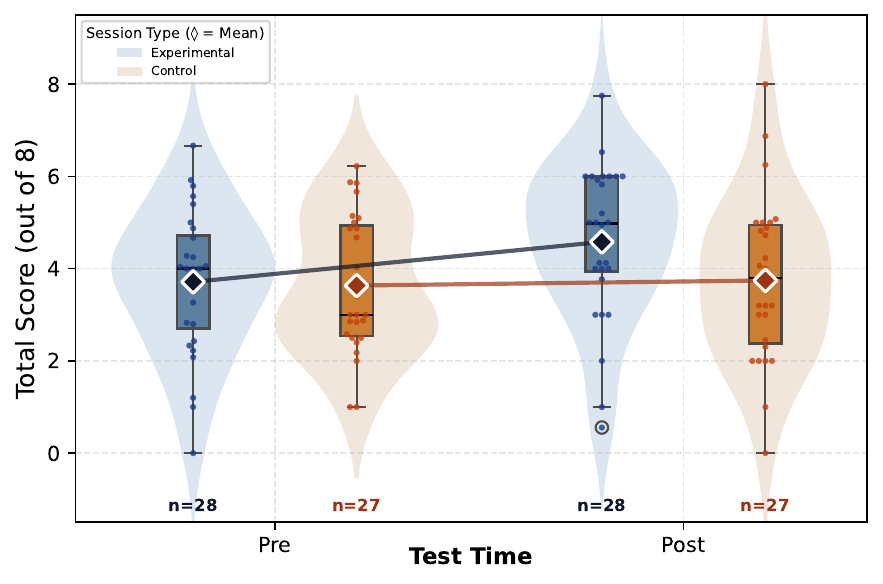}
  \caption{Task performance scores (problems solved, with partial credit) before and after the intervention for experimental and control groups.}
  \label{fig:task-performance}
  \vspace{-2.5mm} 

\end{figure}

The model coefficients and 95\% confidence intervals are reported in Table~\ref{tab:task-performance}. The interaction between Time and Condition suggests that students in the experimental group showed greater gains from pre- to post-test than the control group ($B$ = 0.756, 95\% CI $[-0.020,\ 1.533]$, $p$ = 0.056). There was no significant difference in estimated baseline performance between groups ($B$ = $-0.125$, $p$ = 0.761), indicating no evidence of a baseline difference between conditions. Prior Programming Experience was a significant predictor of performance ($B$ = 0.563, 95\% CI $[0.085, 1.042]$, $p$ = 0.021), estimating that each standard deviation of additional experience corresponded to 0.56 more problems solved. Although not statistically significant at $\alpha$ = 0.05, the effect is practically meaningful: the control group improved minimally (Time: $B$ = 0.110), while the experimental group improved by an estimated 0.87 problems---an effect larger than that of a full standard deviation of prior programming experience.

\begin{table}[!t]
    \centering
    \small  
    \begin{threeparttable}
        \caption{Mixed Linear Model Predicting Task Performance}
        \label{tab:task-performance}
        \begin{tabular}{lrrrr}
            \hline
            \textbf{Parameter} & \textbf{Coef.} & \textbf{Std. Err.} & \textbf{P>|z|} & \textbf{95\% CI} \\
            \hline
            Intercept & 3.760 & 0.293 & 0.000 & [3.185, 4.334] \\
            Condition [Exp.] & -0.125 & 0.411 & 0.761 & [-0.931, 0.681] \\
            Time & 0.110 & 0.283 & 0.688 & [-0.444, 0.664] \\
            Time × Condition & 0.756 & 0.396 & 0.056 & [-0.020, 1.533] \\
            Prior Prog. Experience & 0.563 & 0.244 & 0.021 & [0.085, 1.042] \\
            Self-Efficacy & 0.344 & 0.215 & 0.109 & [-0.077, 0.765] \\
            \hline
            \multicolumn{5}{l}{\textbf{Random Effects}} \\
            Intra-cluster Variance & 1.189 & 0.462 & & \\
            \hline
        \end{tabular}
        \begin{tablenotes}
            \footnotesize
            \item N = 110 observations across 55 participants. Method: REML.
            \item Significance levels: * p < 0.05, ** p < 0.01, *** p < 0.001.
        \end{tablenotes}
    \end{threeparttable}
    \vspace{-2mm} 

\end{table}

\subsection{RQ2: Prompting Self-efficacy}
\label{sec:rq2}

We used the same modeling approach as in RQ1, with prompting self-efficacy (0–10 scale) as the dependent variable. We replaced Self-Efficacy as a covariate with Pre-Test to control for differences in students' initial programming abilities with AI tools. Prior to analysis, we excluded two additional participants due to incomplete survey responses, resulting in 106 observations from 53 students. Model assumptions were met: residuals were approximately normally distributed (Shapiro–Wilk $W \approx 0.98$, $p$ = 0.76), and residual-vs-fitted plots showed no major deviations.

\begin{figure}[ht]
  \centering
  \vspace{-2.5mm} 
  \includegraphics[width=0.85\linewidth]{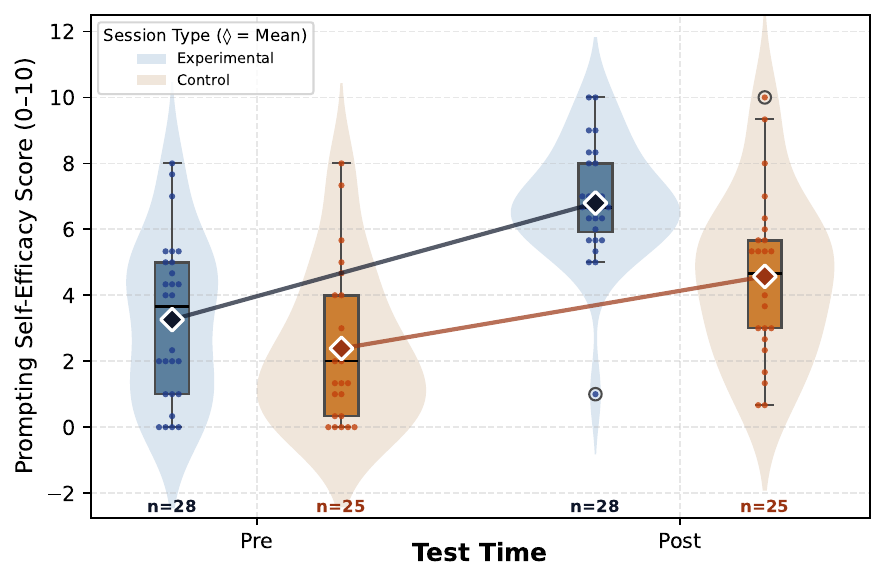}
  \caption{Prompting self-efficacy scores before and after the intervention for experimental and control groups.}
  \label{fig:self-efficacy}
  \vspace{-2.5mm} 
\end{figure}

Table~\ref{tab:self-efficacy} reports the model coefficients and 95\% confidence intervals. The interaction between Time and Condition was statistically significant ($B$ = 1.349, 95\% CI $[0.139,\ 2.559]$, $p$ = 0.029), indicating that students in the experimental group reported significantly higher prompting self-efficacy gains than those in the control group. On average, this corresponds to an increase of approximately 1.35 points (on a 10-point scale) in the experimental condition. This effect is in addition to the main effect of Time that \textit{both} groups experienced ($B$ = 2.187, $p$ < 0.001), indicating that all students reported higher self-efficacy after instruction regardless of condition. Prior Programming Experience was also a significant predictor (see Table~\ref{tab:self-efficacy}), consistent with prior work \cite{ramalingam2004selfefficacy, bui2023prior}. There was no significant baseline difference between conditions.

\begin{table}[!t]
    \centering
    \small  
    \begin{threeparttable}
        \caption{Mixed Linear Model Predicting Prompting Self-Efficacy}
        \label{tab:self-efficacy}
        \begin{tabular}{lrrrr}  
            \hline
            \textbf{Parameter} & \textbf{Coef.} & \textbf{Std. Err.} & \textbf{P>|z|} & \textbf{95\% CI} \\
            \hline
            Intercept & 2.562 & 0.385 & 0.000 & [1.808, 3.316] \\
            Condition [Exp.] & 0.608 & 0.530 & 0.251 & [-0.430, 1.646] \\
            Time & 2.187 & 0.449 & 0.000 & [1.308, 3.066] \\
            Time × Condition & 1.349 & 0.617 & 0.029 & [0.139, 2.559] \\
            Prior Prog. Experience & 1.178 & 0.276 & 0.000 & [0.637, 1.719] \\
            Pre-Test & 0.427 & 0.228 & 0.061 & [-0.020, 0.874] \\
            \hline
            \multicolumn{5}{l}{\textbf{Random Effects}} \\
            Intra-cluster Variance & 1.153 & 0.430 & & \\
            \hline
        \end{tabular}
        \begin{tablenotes}
            \footnotesize
 
            \item N = 106 observations across 53 participants. Method: REML.
            \item Significance levels: * p < 0.05, ** p < 0.01, *** p < 0.001.
        \end{tablenotes}
    \end{threeparttable}
    \vspace{-3.5mm} 
\end{table}

\section{Discussion}

Our findings suggest that a brief, targeted intervention may produce meaningful, but statistically uncertain gains in prompt-based programming. On average, the experimental group improved by +10.8 percentage points on the test and +35.4 points in self-efficacy, compared to +1.1 and +21.9, respectively, in the control group. While the effect on task performance was not statistically significant ($p$ = 0.056), the estimated magnitude was comparable to that of a full standard deviation of prior programming experience, suggesting that instruction may help close the gap between learners with and without formal CS backgrounds. Still, the magnitude of the effect (less than one additional problem solved on average) indicates that while 45 minutes of instruction may produce meaningful gains, students clearly did not achieve mastery, and further instruction is likely needed. Our results suggest a likely range of impacts that a short intervention like ours would have on learning and because this range includes zero (95\% CI $[-0.020,\ 1.533]$), future research with a larger population is needed to confirm the effectiveness of the intervention. 

It is useful to know whether a short lesson on prompting can produce meaningful gains. Prompting is a new and understudied skill, and instructors today are either dedicating significant time to it or skipping it entirely. Our results help establish the value of something in between. At the same time, brief interventions can fail, and prompting may require more practice than 45 minutes can provide. Because the lesson and its guided practice are intertwined, part of the experimental group's advantage may reflect familiarity with the assessment format rather than prompting skills alone, since the control group did not practice in this format. Separating these effects is a useful direction for future work, but addresses a different research question than the one we ask here.


One explanation for the observed effects comes from the Gulf of Envisioning framework \cite{subramonyam2024bridging}. By teaching students to explicitly articulate inputs, outputs, and transformations using a structured template, we addressed the information content problem, which recent work identifies as more predictive of prompt success than vocabulary alone \cite{lucchetti2025substance}. This aligns with recent work on the ROPE intervention \cite{ma2025rope}, which found that structured instruction can improve prompting outcomes, and with Vadaparty et al.'s CS1-LLM course \cite{vadaparty2024cs1llm}, where teaching prompt engineering alongside foundational concepts like problem decomposition helped students tackle more complex projects than a traditional CS1 typically allows.

However, the modest and uncertain effect size suggests important limitations. One possibility is that students learned the surface structure of prompting without fully internalizing how to express precise computational logic. The 45-minute duration may have been sufficient to introduce the template but insufficient for students to develop fluency. Additionally, the control condition was designed as a business-as-usual baseline rather than a no-instruction condition. Code comprehension skills may have helped students interpret LLM-generated code on Promptly, which narrows the gap between conditions.


\textbf{Implications for teaching}: Our intervention suggests that prompt-based programming can blend foundational CS0 concepts (e.g., function inputs/outputs) with more advanced practices (e.g., test-driven development, edge case reasoning) and prompt-specific strategies. Rather than replacing traditional coding instruction, prompt-based programming may offer a complementary pathway, particularly for students who are intimidated by syntax or control flow. For these learners, writing structured prompts may provide a more accessible entry point into computing. However, our results showed that while the performance gains were meaningful, they were also modest (less than one additional problem solved, on average). Sustained practice would likely be needed for learners to achieve fluency in prompt-based programming tasks.

\textbf{Implications for research}: These limited gains raise key questions about the boundaries of current instructional approaches. Prior studies suggest that learners often hold incomplete or inaccurate mental models of how LLMs interpret prompts \cite{nguyen2024how}, and our findings support the need to examine how those models evolve, or persist, even after instruction. Future work should analyze student-generated prompts to better understand where students succeed, where they struggle, and what misconceptions remain unaddressed. While our brief intervention shows promise for simple, well-scoped tasks, it may not generalize easily to the kinds of open-ended, multi-step problems common in real-world end-user programming. As programs grow in complexity, natural language alone may not suffice, especially for debugging or coordinating across multiple components \cite{wu2022promptchainer}. Future research should explore how prompt-based instruction can be extended or hybridized (e.g., with editing tools or code visualization) to better support complex programming goals.

\vspace{-1mm}
\section{Limitation}
\label{sec:limitations}

While our findings highlight the promise of prompt-based instruction, we note four limitations. First, the 45-minute intervention is short relative to typical programming courses, though this represents a single class period and was a deliberate design choice for our target population (see Discussion). Second, all sessions were led by the first author, who also designed the curriculum, introducing potential experimenter expectancy bias. Although we used structured slides and facilitator notes to ensure consistency, differences in delivery or enthusiasm may have influenced outcomes. Third, sessions were conducted in both in-person and online formats. While delivery format may influence learning outcomes, modality was distributed comparably across conditions (experimental: 2 in-person, 5 online; control: 2 in-person, 6 online), so any such effects would not systematically favor either group. Fourth, we focused on short-term gains; longitudinal research is needed to assess whether prompting skills persist and transfer. Additionally, we did not analyze prompt content, limiting our ability to explain why the intervention was effective. In the future, we plan to analyze student prompts to identify which strategies students applied.

\vspace{-1mm}
\section{Conclusion}
For many end-user programmers, the goal is not to write perfect code but to articulate intentions clearly enough for an AI to generate it. Rather than replacing traditional programming, prompt-based programming offers a complementary path for those learning to collaborate with AI. Our findings suggest that even a short, lightweight intervention can improve students' performance and self-efficacy in prompt-based programming.



\bibliographystyle{ACM-Reference-Format}
\bibliography{main}

\end{document}